\documentclass[unsortedaddress,prfluids]{revtex4-2}
\usepackage[T1]{fontenc}
\usepackage[utf8]{inputenc}
\usepackage{lmodern}
\usepackage[a4paper]{geometry}
\usepackage{amsmath, esint, amssymb}
\usepackage{amsfonts}
\usepackage{float}
\usepackage{graphicx}
\usepackage{epstopdf}
\usepackage{xcolor}
\usepackage{tabularx}
\usepackage[hyperpageref]{backref}  
\usepackage{listings}
\usepackage{enumitem}
\usepackage{subcaption}

\usepackage[normalem]{ulem}

\definecolor{darkred}{RGB}{122,0,0}
\definecolor{darkyellow}{RGB}{122,122,0}
\definecolor{darkgreen}{RGB}{0,122,0}
\definecolor{cyan}{RGB}{0,122,122}
\definecolor{darkblue}{RGB}{0,0,122}
\definecolor{darkviolet}{RGB}{122,0,122}
\definecolor{darkgray}{RGB}{122,122,122}
\definecolor{Prune}{RGB}{99,0,60}

\usepackage[colorlinks=true,
linkcolor= darkred,
citecolor= darkblue,
urlcolor= darkblue,
hyperfigures=false
breaklinks=true]{hyperref}

\definecolor{definitioncolor}{rgb}{0.5,0,0} 
\definecolor{postulatecolor}{rgb}{0,0,0.5}
\definecolor{theoremcolor}{rgb}{0,0.5,0} 
\definecolor{examplecolor}{rgb}{0.0,0.0,0.0}

\usepackage{array}
\newcolumntype{L}[1]{>{\raggedright\let\newline\\\arraybackslash\hspace{0pt}}m{#1}}
\newcolumntype{C}[1]{>{\centering\let\newline\\\arraybackslash\hspace{0pt}}m{#1}}
\newcolumntype{R}[1]{>{\raggedleft\let\newline\\\arraybackslash\hspace{0pt}}m{#1}}

\newcommand{\lra}[1]{\langle #1 \rangle }

\begin{document}

\title{\bf  Heat-flux Fluctuations reveals regime transitions in Rayleigh-Bénard convection}
\author{Vincent Labarre$^1$} 
\author{St\'ephan Fauve$^2$}
\author{Sergio Chibbaro$^{3,4}$}
\affiliation{$^1$ Sorbonne Universit\'e, CNRS, UMR 7190, Institut Jean Le Rond d'Alembert, F-75005 Paris, France}
\affiliation{$^2$ Laboratoire de Physique de l'Ecole Normale Sup\'erieure, CNRS, PSL Research University,
Sorbonne Universit\'e, Universit\'e Paris Cit\'e, F-75005 Paris, France}
\affiliation{$^3$ Universit\'e Paris-Saclay, CNRS, UMR 9015, LISN, F - 91405 Orsay cedex, France}
\affiliation{$^{4}$ SPEC, CNRS UMR 3680, Université Paris-Saclay, CEA Saclay, Gif-sur-Yvette, France}

\begin{abstract}
The study of the transitions among different regimes in thermal convection has been an issue of paramount importance in fluid mechanics.
While the bifurcations at low Rayleigh number, when the flow is laminar or moderately chaotic, have been fully understood for a long time, transitions at higher Rayleigh number are much more difficult to be clearly identified. Here, through a numerical study of the two-dimensional Rayleigh-B\'enard convection covering four decades in Rayleigh number for two different Prandtl numbers, we find a clear-cut transition by considering the fluctuations of the heat flux through a horizontal plane, rather than its mean value.
More specifically, we have found that this sharp transition is displayed by a jump of the ratio of the root-mean-square fluctuations of the heat flux to its mean value and occurs at $Ra/Pr \approx 10^9$.
Above the transition, this ratio is found to be constant in all regions of the flow, while taking different values in the bulk and at the boundaries.
Below the transition instead, different behaviors are observed at the boundaries and in the bulk: at the boundaries, this ratio decreases with respect to the Rayleigh number whereas it is found to be constant in the bulk for all values of the Rayleigh number.
Through this numerical evidence and an analytical reasoning we confirm what was already observed in experiments, that is the decrease of the ratio of root-mean-square fluctuations of the heat flux to its mean value, observed at the boundaries below the transition, can understood in terms of the law of large numbers.
\end{abstract}
\maketitle
	
	

\section{Introduction \label{section1}}

Thermal convection is related to flows generated by the buoyancy force that results from temperature gradients within a fluid \cite{landau2013fluid}. 
They are widely observed in the atmosphere,
in the Earth mantle or in its outer liquid core as well as in the core of many planets and stars. Thermal convection also plays an important role in many industrial processes.
A quantity of primary interest is the heat flux transported by a convective flow, for instance if one has to evaluate how is evacuated the heat generated in the inner core of a star or the cooling efficiency of a flow in some industrial process \cite{ahlers2009heat,verma2018physics}. 

This problem still involves open questions, even in one of the simplest configuration of a convective flow, the so-called Rayleigh-B\'enard convection~\cite{manneville2006, ahlers2009heat,lohse2010small,chilla2012new,xia2013current}. It consists of studying the problem for a horizontal layer of fluid of height $H$ heated from below and therefore submitted to a temperature difference $\Delta T$. In the  Oberbeck-Boussinesq approximation
\cite{spiegel1960,mihaljan1962} the problem involves two dimensionless parameters, the Rayleigh number, $Ra = g \beta H^3 \Delta T / \nu \kappa$, where $g$ is the acceleration of gravity, $\beta$ is the volumetric  thermal expansion coefficient, $\nu$ is the kinematic viscosity and $\kappa$ is the thermal diffusivity, and the Prandtl number, $Pr = \nu/\kappa$.  If the layer has a finite horizontal extent $L$, one has also to take into account the aspect ratio $\Gamma = L/ H$. 
In dimensionless form, the heat flux is described by the Nusselt number, $Nu$, which is the ratio of the convective heat flux to the one that would exist in the absence of convection for the same $\Delta T$. If one discards the aspect ratio, dimensional analysis implies that $Nu = f(Ra, Pr)$. The determination of the function $f$ is out of reach of any analytical calculation except in the vicinity of the convection threshold \cite{schluter1965}. 
Further above the convection threshold, experimental measurements of the heat flux have been fitted by laws of the form $Nu \propto Ra^{\alpha} Pr^{\gamma}$ with $\alpha = 1/5, 1/4, 2/7, 1/3, 1/2$ to quote some of them. Simple laminar flow models, dimensional analysis or ad-hoc arguments have been used to try to justify them as reviewed in \cite{kraichnan1962,spiegel1971convection,siggia1994high}. 

The validity of most of these laws is of course limited to a finite range of $Ra$ and $Pr$.  Some of them have been recovered later
and the crossover between neighbouring regimes has been calculated by considering kinetic and thermal dissipation and determining whether the bulk or the boundary layer contribution is the dominant one \cite{grossmann2000scaling}. However, the main problem that has been considered more than half a century ago is whether $f$ could become a power law in the limit of strongly developed convection, $Ra \rightarrow \infty$. This would mean that there exists only one relevant dimensionless parameter in that limit. A first proposal has been made by Malkus \cite{malkus1954heat,malkus1954discrete} using the assumption that turbulent convection maximises the heat flux and expanding the motion in a sum of linear modes truncated such that the highest mode is neutral.  He thus found $\alpha = 1/3$. This also can be recovered assuming that the thermal boundary layers are at marginal stability with respect to convection. More generally, this scaling law is obtained if the heat flux is determined locally by the structure of the thermal boundary layer such that the height $H$ of the layer can be discarded \cite{priestley1954}. A different answer is based on the belief in turbulence theory which assumes that macroscopic transport properties do not depend any more on microscopic transport coefficient when turbulence is fully developed. Discarding $\nu$ and $\kappa$ gives $\alpha = \gamma = 1/2$ \cite{kraichnan1962,spiegel1971convection,grossmann2000scaling}. 
While a heuristic rationale to systemise the data has been proposed, the arguments leading to both scaling laws are equally convincing or questionable. Experimental results do not provide a clear picture~\cite{chavanne2001turbulent,niemela2003,he2012transition} and there is some debate on data analysis \cite{doering2019absence,roche2020}.
Generally speaking, to sharply capture transitions in such kind of turbulent flows is difficult, notably at high Rayleigh numbers.
Beside possible experimental issues, the transitions are usually related to a change of scaling in the relation between $Nu$ and $Ra$, and yet the scaling exponents are often nearby and the change in slope should be observed over many decades to be convincing.

On the other hand, the importance of fluctuations on the mean heat transfer has been emphasized long ago \cite{howard1966}. Most notably, they have been considered in relation to the small-scale properties of turbulent convection~\cite{lohse2010small}.
Fluctuations in the form of thermal plumes have been taken into account in a dimensional argument to find  $\alpha = 2/7$ \cite{castaing1989scaling,procaccia1991transitions}. 
They have been also considered~\cite{grossmann2004fluctuations} to revisit earlier predictions~\cite{grossmann2000scaling}.  
Other few works have looked at the global probability density function of different observables~\cite{zhang2017statistics, kaczorowski2013turbulent, shishkina2007local,zonta2016entropy} in relation to non-equilibrium statistical mechanics.
Finally, a series of works have taken into account fluctuations in the determination of the structure of boundary layers \cite{shishkina2015thermal,shishkina2017mean,tai2021heat}. 
Although in those works the authors took into account the existence of fluctuations in order to determine how they could affect the mean heat flux, they did not study the characteristics of the fluctuations of the heat flux and their possible scaling laws. 
Therefore, despite the general relevance of fluctuations, almost no work has considered heat-flux fluctuations in a systematic way with regard to the scaling laws.

It should be emphasized that in the limit of an infinite aspect ratio, it has been usually assumed that the heat flux does not fluctuate in time. The very definition of the Nusselt number indeed assumes that the spatially averaged temperature on any horizontal plane of infinite extent is constant. It is not clear that this is a correct assumption for the realistic case of finite, even though large aspect ratio. This relies on the fact that no coherent large scale flow would exist for large enough aspect ratio which does not seem to be true. 
It has been indeed found that the total heat flux at the horizontal boundaries display fairly large fluctuations \cite{aumaitre2003statistical}. 
For Rayleigh numbers in the range $10^7 < Ra < 10^9$, 
experimental evidence indicates that the root mean square (rms) of the heat-flux fluctuations is proportional to $\Delta T$, which corresponds to a ratio of the rms of fluctuations to the mean heat-flux that decreases like $Ra^{-\delta}$ with $\delta \approx \alpha$. 
This scaling law can be understood using the law of large numbers for the fluctuations of the thermal boundary layer. 
Moreover, a lot of convection experiments are operated by applying a constant heat flux to the bottom plate instead of maintaining its temperature constant. Fluctuations of the total heat flux therefore generate fluctuations of the bottom plate temperature. It has been observed that these fluctuations strongly increase above $Ra \sim 10^{12}$ \cite{gauthier2008evidence}, and this increase has been related to a boundary layer transition.

The purpose of the present work is to analyse the behavior of heat-flux fluctuations at varying the Rayleigh and Prandtl number.
We want to understand whether the fluctuations may be a key observable to capture the physics of the turbulent convection, and notably possible bifurcations between flow states.
Furthermore, we want to verify if the predictions made in previous experiments~\cite{aumaitre2003statistical} can be confirmed and are robust.

To achieve our goal, we have performed resolved numerical simulations (DNS) of two-dimensional (2D) Rayleigh-B\'enard turbulence over a large span of Rayleigh numbers and considering two Prandtl fluids, namely $Pr=0.71$ (air) and $Pr=7$ (water).
While real-world applications are three-dimensional, in fact numerical simulations in three dimensions (3D) are computationally exorbitant~\cite{stevens2010radial,shishkina2010boundary},
forbidding large parametric studies.
In addition, theoretical analysis are based on 2D boundary-layer models~\cite{castaing1989scaling,siggia1994high,grossmann2000scaling,shishkina2015thermal}, or assumptions that apply to 3D as well to 2D flows.
Moreover, even considering Rayleigh numbers not too high, 
to have results well converged in statistics of higher order than the first moment 
is practically impossible.
Moreover, while 2D Rayleigh-B\'enard simulations are different from the 3D with regard to integral observables at small Pr~\cite{schmalzl2004validity}, yet they reproduce most of 3D features~\cite{van2013comparison}, and may have an interest \emph{per se}~\cite{castillo2016reversal}.
In particular, it has been shown that 2D and 3D Rayleigh-B\'enard are very similar at Prandtl numbers higher than 1.
At Prandtl numbers around or inferior to $1$,
differences may be more important, and 
 the integral quantities and flow state in 2D have a stronger dependence on the aspect-ratio $\Gamma$ than in 3D.
 Yet, while differences are significant at low Rayleigh numbers $Ra\lesssim 10^6$ and at small aspect ratio $\Gamma \approx 0.5$, results in 2D are again in good agreement at higher $Ra$ with $\Gamma \approx 1$~\cite{van2013comparison}.

In the last years, the 2D configuration has been therefore widely used for parametric analysis in a variety of configurations, to get new insights and test theories~\cite{huang2013counter,hewitt2012ultimate,zhang2017statistics,zhu2018transition,wang2021regime}.
In the present work, we shall compare our 2D numerical results with the experimental observations obtained in the same range of parameters to 
further verify to which extent our findings may be applied to the 3D case. On the other hand, the numerical simulations will permit to access a range of Rayleigh numbers which were not possible to be investigated in the experiments.

DNS have the advantage of avoiding artefacts due to experimental issues, thus constituting actually idealised experiments. 
Another advantage is that they permit to access to all kind of details of the flow field,
most of which are unavailable in experiments.
For instance, it was possible experimentally only to measure fluctuations at boundaries~\cite{aumaitre2003statistical}, whereas numerically we are able to have also data about the core of the flow.
The level of details provided by DNS is then helpful to get physical insights on the complex phenomena, and specifically on the bifurcations. 

This paper is organized as follows: in the next section, 
we briefly present the model and provide the characteristics of the numerical simulations. 
The section \ref{section3} is focused on the results.
Firstly, in section \ref{section3a}, we give some information about the global dynamics. This part is mainly meant to show that the present results are consistent with previous studies of Rayleigh-B\'enard convection, and most notably in two dimensions, while the focus of the work is on the fluctuations.
 Since the case at $Pr=7$ has been much less analysed, this section allows to remind the main differences with the case at $Pr\approx1$. \\
 In section \ref{section3b},
 we present the first results about the statistics of the heat flux.
 We discuss the standard relation between the mean Nusselt $\lra{Nu}$ and the Rayleigh number,
 which still displays a scaling consistent with previous studies and does not show a clear transition 
 between different scalings. Then, we present the probability density function of the fluctuating 
 Nusselt number at the wall and at the mid-plane, which display a mild dependence with Rayleigh. 
An important information given by these statistics is that the pdfs in the present work are in good agreement with the only experiments available. That corroborates the fact that present two-dimensional simulations are also relevant for three dimensional flows.  \\
In section \ref{section3c}, we present the main result of the work.
Specifically, we show the root mean square of the fluctuating heat-flux as a function of the ratio $Ra/Pr$, at the walls and at the mid-plane. This statistics displays a neat transition 
at $Ra/Pr=10^9$ in all cases. In particular, at the walls the heat-flux decreases with $Ra$ before the transition, and displays a plateau after. We give also an analytical explanation of the behavior. We make a final discussion of the results and conclude in section \ref{section4}.

\section{Theoretical and numerical model \label{section2}}

\begin{figure}[h]
	\begin{center}
		\includegraphics[scale=1.0]{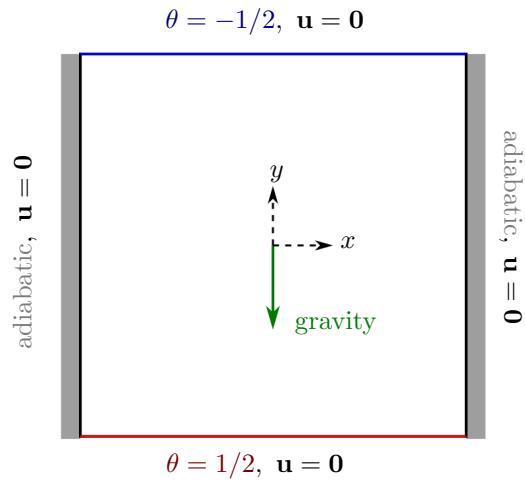}
		\caption{Figure of the square cell used in this study. $x$ and $y$ denote respectively the horizontal coordinate and the vertical coordinate. The origin is placed at the center of the cell. The domain is therefore given by $x \in [-1/2:1/2], y \in [-1/2:1/2]$. $\theta$ is the dimensionless temperature and $\textbf{u}$ is the dimensionless velocity. Constant temperatures are imposed on top and bottom boundaries, and adiabatic conditions are imposed on lateral boundaries. No-slip boundary conditions are used. }\label{fig1}
\end{center}
\end{figure} 
In this study, we consider a 2D fluid contained in a square cell heated from below and cooled form above where the bottom and top plates are orthogonal to the uniform gravitational field (Fig. \ref{fig1}).

The problem is mathematically described in the Oberbeck-Boussinesq approximation, so that the evolution equations for the dimensionless velocity $\textbf{u} = (u_x, u_y)$ and the dimensionless temperature $\theta$ read as:
\begin{align}
\label{eq:Continuity}
\dfrac{\partial u_x}{\partial x} + \dfrac{\partial u_y}{\partial y} &= 0 , \\
\label{eq:Impulsionx}
\dfrac{\partial u_x}{\partial t} + u_x \dfrac{\partial u_x}{\partial x} + u_y \dfrac{\partial u_x}{\partial y} &= - \frac{\partial p}{\partial x} + \sqrt{\frac{Pr}{Ra}} \left( \frac{\partial^2 u_x}{\partial x^2} + \frac{\partial^2 u_x}{\partial y^2} \right), \\
\label{eq:Impulsionz}
\dfrac{\partial u_y}{\partial t} + u_x \dfrac{\partial u_y}{\partial x} + u_y \dfrac{\partial u_y}{\partial y} &= \theta  - \frac{\partial p}{\partial y} + \sqrt{\frac{Pr}{Ra}} \left( \frac{\partial^2 u_y}{\partial x^2} + \frac{\partial^2 u_y}{\partial y^2} \right), \\
\label{eq:Heat}
\dfrac{\partial \theta}{\partial t} + u_x \dfrac{\partial \theta}{\partial x} + u_y \dfrac{\partial \theta}{\partial y} &= \frac{1}{\sqrt{Ra Pr}} \left( \frac{\partial^2 \theta}{\partial x^2} + \frac{\partial^2 \theta}{\partial y^2} \right),
\end{align}

The free fall velocity $\sqrt{\beta \Delta T g H}$ has been used to make the velocity dimensionless.
 
We have applied no-slip boundary conditions everywhere for the velocity, adiabatic conditions for the temperature on the lateral boundaries, and constant temperature on the top and bottom boundaries, as specified  in the Fig. \ref{fig1}.
\begin{table*}
	\begin{center}
				\begin{tabular}{@{}|C{1.5cm}|C{1.5cm}|C{1.5cm}|C{1.5cm}|C{1.5cm}|C{1.5cm}|C{1.5cm}|}
			\hline
			\textbf{Run \#}  & $\boldsymbol{Ra}$ & $\boldsymbol{Pr}$ & $\textbf{N}$ & $\boldsymbol{T}$ & $\boldsymbol{\Lambda_u}$ & $\boldsymbol{\Lambda_T}$ \\
			\hline
			$1$ & $1e7$ & $0.71$ & $256$ & $2 000$ & $-0.008$ & $-0.010$ \\
			$2$ & $2e7$ & $0.71$ & $256$ & $1 400$ & $-0.005$ & $-0.012$ \\
			$3$ & $5e7$ & $0.71$ & $256$ & $1 900$ & $-0.030$ & $-0.034$ \\
			$4$ & $1e8$ & $0.71$ & $256$ & $191 641$ & $-0.020$ & $-0.039$ \\
			$5$ & $1e8$ & $0.71$ & $512$ & $2 300$ & $0.045$ & $0.005$ \\
			$6$ & $1e8$ & $0.71$ & $1024$ & $2 000$ & $0.058$ & $0.010$ \\
			$7$ & $2e8$ & $0.71$ & $512$ & $2 000$ & $0.022$ & $-0.005$ \\
			$8$ & $5e8$ & $0.71$ & $1024$ & $1 210$ & $0.021$ & $-0.009$ \\
			$9$ & $1e9$ & $0.71$ & $1024$ & $3 000$ & $0.007$ & $-0.026$ \\
			$10$ & $2e9$ & $0.71$ & $2048$ & $1 673$ & $-0.126$ & $-0.019$ \\
			$11$ & $5e9$ & $0.71$ & $2048$ & $2 000$ & $-0.019$ & $0.047$ \\
			$12$ & $1e10$ & $0.71$ & $2048$ & $2 000$ & $-0.157$ & $-0.030$ \\
			$13$ & $1e11$ & $0.71$ & $2048$ & $1748$ & $-0.057$ & $-0.085$ \\
			$14$ & $1e7$ & $7$ & $256$ & $2 500$ & $-0.008$ & $-0.012$ \\
			$15$ & $2e7$ & $7$ & $256$ & $3 000$ & $-0.003$ & $-0.014$ \\
			$16$ & $5e7$ & $7$ & $512$ & $1 784$ & $0.0018$ & $-0.0066$ \\
			$17$ & $1e8$ & $7$ & $256$ & $1 300$ & $-0.016$ & $-0.048$ \\
			$18$ & $1e8$ & $7$ & $512$ & $2 000$ & $-0.005$ & $-0.014$ \\
			$19$ & $2e8$ & $7$ & $512$ & $3 000$ & $-0.012$ & $-0.018$ \\
			$20$ & $5e8$ & $7$ & $1024$ & $1 286$ & $0.003$ & $-0.008$ \\
			$21$ & $1e9$ & $7$ & $1024$ & $1 500$ & $-0.007$ & $-0.012$ \\
			$22$ & $2e9$ & $7$ & $2048$ & $574$ & $0.00012$ & $-0.0068$ \\
			$23$ & $5e9$ & $7$ & $2048$ & $413$ & $0.012$ & $-0.016$ \\
			$24$ & $1e10$ & $7$ & $2048$ & $714$ & $-0.028$ & $-0.0005$ \\
			$25$ & $2e10$ & $7$ & $2048$ & $400$ & $-0.033$ & $-0.026$ \\
			$26$ & $1e11$ & $7$ & $2048$ & $350$ & $0.061$ & $-0.087$ \\
			\hline
		\end{tabular}
		\caption{\label{tab1} Summary of the different direct numerical simulations. The uniform cartesian mesh is composed by $N \times N$ nodes. The size of the time interval in the statistically steady state is $T$. $\Lambda_u$ and $\Lambda_T$ are the relative errors estimated by computing the Nusselt number using its definition or the kinetic and thermal dissipations, Eq. (\ref{eq:Lambda}).}
	\end{center}	
\end{table*}

 The equations (\ref{eq:Continuity})-(\ref{eq:Heat}), together with the boundary conditions have been solved for different $Ra$ and $Pr$ using the open-source code Basilisk \cite{Basilisk}.
Basilisk uses Finite-Volume numerical schemes, notably  
with Bell-Colella-Glaz advection scheme \cite{bell1989}, and a pressure-correction scheme for the velocity-pressure coupling, with a global second-order precision. 
The code has been now comprehensively validated in turbulent flows, and most notably in Rayleigh-B\'enard convection~\cite{castillo2016reversal,castillo2017turbulent, castillo2019cessation,valori2020weak}. 
All simulations have been performed on a uniform Cartesian grid, and with a variable time-step that verifies the condition CFL < 0.5. 
The mesh size has been chosen to fulfil with the standard criteria provided to well resolve all the boundary layers~\cite{stevens2010radial,shishkina2010boundary}.
The details of the numerical simulations are presented in the Table \ref{tab1}.

For the purpose of this work, the key quantity of the system is the instantaneous vertical heat-flux, whose dimensionless definition is 
\begin{equation}
Nu({\bf x},t) = \sqrt{Ra Pr} ~ \theta ~ u_y - \frac{\partial \theta}{\partial y},
\end{equation}
this quantity represents thus an instantaneous Nusselt number, whereas the mean heat-flux is given by the Nusselt number defined as $\lra{Nu}$.

As customary in Rayleigh B\'enard convection, 
in order to assess the resolution of the numerical method, and the statistical convergence we have used the consistency relation for the mean heat transfer \citep{shraiman1990heat,siggia1994high,verzicco2003numerical}: 
\begin{equation}
\lra{Nu}\equiv1+\sqrt{Ra Pr} \lra{u_y \theta}=Nu_{\epsilon}\equiv 1+\sqrt{Ra Pr} \lra{\epsilon}= Nu_{\epsilon_T}\equiv \sqrt{Ra Pr} \lra{\epsilon_T},
\label{eq:consistency}
\end{equation} 
where $ \lra{}$ indicates statistical averaging, 
$\epsilon$ is the kinetic energy dissipation-rate and $\epsilon_T$ is the temperature-variance dissipation rate.
Actually, we have computed the following spatial-averaged quantities at each time
\begin{align}
Nu_g(t) &\equiv \int\limits_{-1/2}^{1/2} ~ \int\limits_{-1/2}^{1/2} ~ Nu(x,y,t) ~ \mathrm{d}x ~ \mathrm{d}y, \\
\sqrt{Ra Pr} ~ \epsilon(t) + 1 &\equiv Pr ~ \int\limits_{-1/2}^{1/2} ~ \int\limits_{-1/2}^{1/2} ~ \dfrac{1}{2} \left[ \boldsymbol{\nabla} \textbf{u} + (\boldsymbol{\nabla} \textbf{u})^T \right]^2 ~ \mathrm{d}x ~ \mathrm{d}y + 1, \\
\sqrt{Ra Pr} ~ \epsilon_T(t) &\equiv \int\limits_{-1/2}^{1/2} ~ \int\limits_{-1/2}^{1/2} ~ \left( \boldsymbol{\nabla} T \right)^2 ~ \mathrm{d}x ~ \mathrm{d}y, 
\end{align}
and then averaged over time to get the relation (\ref{eq:consistency}). These relations clarify also the dimensionless definitions of the dissipation rates $\epsilon, \epsilon_T$.

Specifically, we have quantified the accuracy and consistency of our simulations by computing the following relative errors~\cite{scheel2013resolving}
\begin{equation}
\label{eq:Lambda}
\Lambda_u = \dfrac{\sqrt{Ra Pr} \langle \epsilon \rangle - \left( \langle Nu_g \rangle - 1 \right)}{\langle Nu_g \rangle -1},~~~
\Lambda_T = \dfrac{\sqrt{Ra Pr} \langle \epsilon_T \rangle - \langle Nu_g \rangle}{\langle Nu_g \rangle}.
\end{equation}  
As shown in Table \ref{tab1}, $\Lambda_u$ and $\Lambda_t$ are of few $\%$ and in most cases less than $1\%$, even  
averaging over just $15$ convective times. 

All the runs are well resolved, possibly for the runs at $Ra=1e11$ a longer time-averaging would improve convergence. 

In this study, we are interested in the statistics of the following observables
\begin{align}
Nu_b(t) &\equiv \int\limits_{-1/2}^{1/2} ~ Nu(x,-1/2,t) ~ \mathrm{d}x, \\  
Nu_m(t) &\equiv \int\limits_{-1/2}^{1/2} ~ Nu(x,0,t) ~ \mathrm{d}x, \\  
Nu_t(t) &\equiv \int\limits_{-1/2}^{1/2} ~ Nu(x,1/2,t) ~ \mathrm{d}x,   
\end{align}
which are respectively the Nusselt number intergrated over the bottom boundary, the Nusselt number intergrated over the middle line, and the Nusselt number intergrated over the upper boundary. 
 
The root mean square (rms) of $Nu_b$, $Nu_m$ and $Nu_t$ are:
\begin{equation}
\sigma_b \equiv \sqrt{\left\langle ( Nu_b - \langle Nu_b \rangle )^2 \right\rangle}~,~
\sigma_m \equiv \sqrt{\left\langle ( Nu_m - \langle Nu_m \rangle )^2 \right\rangle}~,~
\sigma_t \equiv \sqrt{\left\langle ( Nu_t - \langle Nu_t \rangle )^2 \right\rangle}
\label{eq:sigma} 
\end{equation}

where $\langle \cdot \rangle$ represents the temporal average in statistically steady state.
To further assess the accuracy of the numerical approach and notably the statistical convergence
of the main observables, we have computed the average of the three different Nusselt numbers 
$Nu_b$, $Nu_m$ and $Nu_t$, and we have found them indistinguishable, as shown in Table \ref{tab2} in appendix.

\begin{figure}[h]
	\begin{center}
		\includegraphics[scale=0.8]{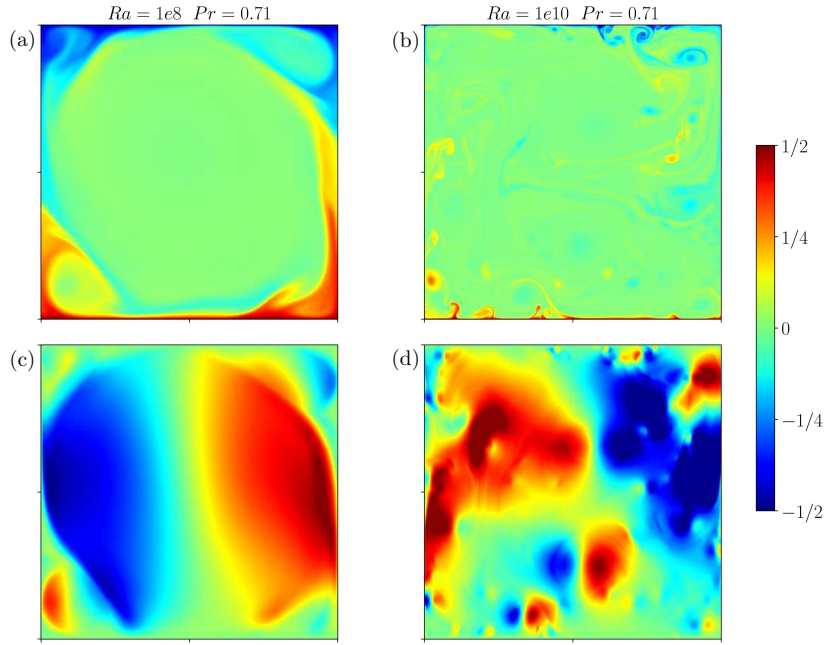}
		\caption{Snapshots of the normalized temperature and vertical velocity for two simulations with $Pr=0.71$. (a) Temperature field for $Ra=10^8$. (b) Temperature field for $Ra=10^{10}$. (c) Vertical velocity field for $Ra=10^8$. (d) Vertical velocity field for $Ra=10^{10}$. \label{Fig:TemperatureVelocity}}
	\end{center}
\end{figure}
Since in the present work we compare our numerical results to previous experiments~\cite{aumaitre2003statistical}, we give some short information about the experimental set-up, while the details are to found in the original paper~\cite{aumaitre2003statistical}.
The authors used different experimental devices. 
The first set-up was a cylindric cell of aspect ratio $\Gamma \equiv D/d = 1$ where $D$ is the diameter of the cell and $d$ its height. The second set-up was a cubic cell. The third set-up was a cylindric cell of aspect ratio $\Gamma = 1/2$. They used two fluids: water ($Pr \simeq 7$) and mercury ($Pr \simeq 0.02$). The measurements of the heat-flux was done over the bottom plate for the cubic cell. For cylindric cells, the heat-flux was measured over a region of the bottom plate which was large compared to the boundary layer thickness.

\section{Results \label{section3}}
\subsection{Qualitative observations}
\label{section3a}
\begin{figure}[h]
	\begin{center}
		\includegraphics[scale=0.5]{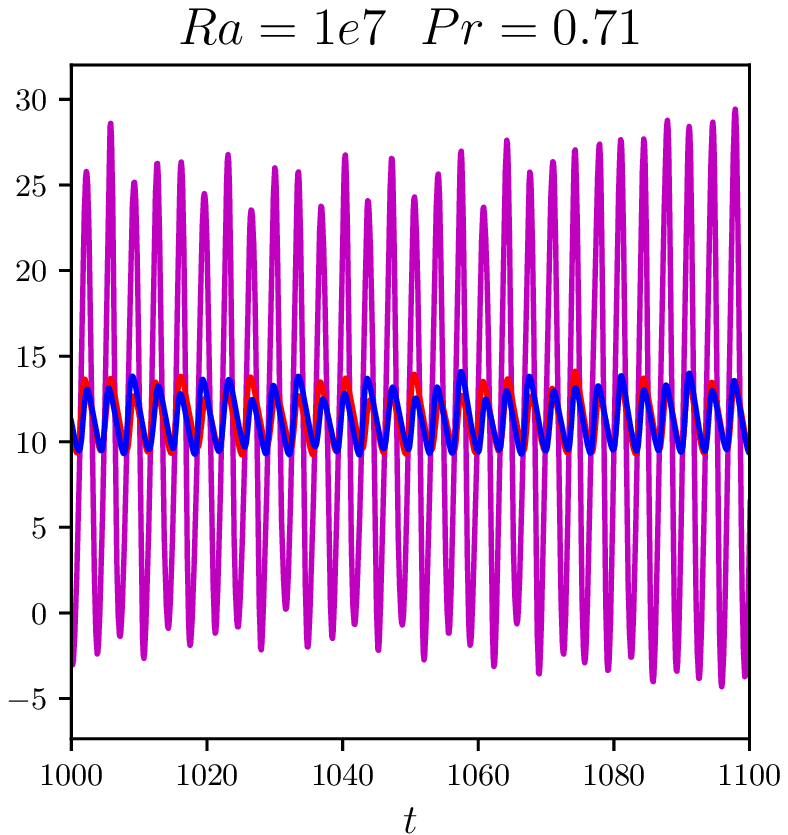}
		\includegraphics[scale=0.5]{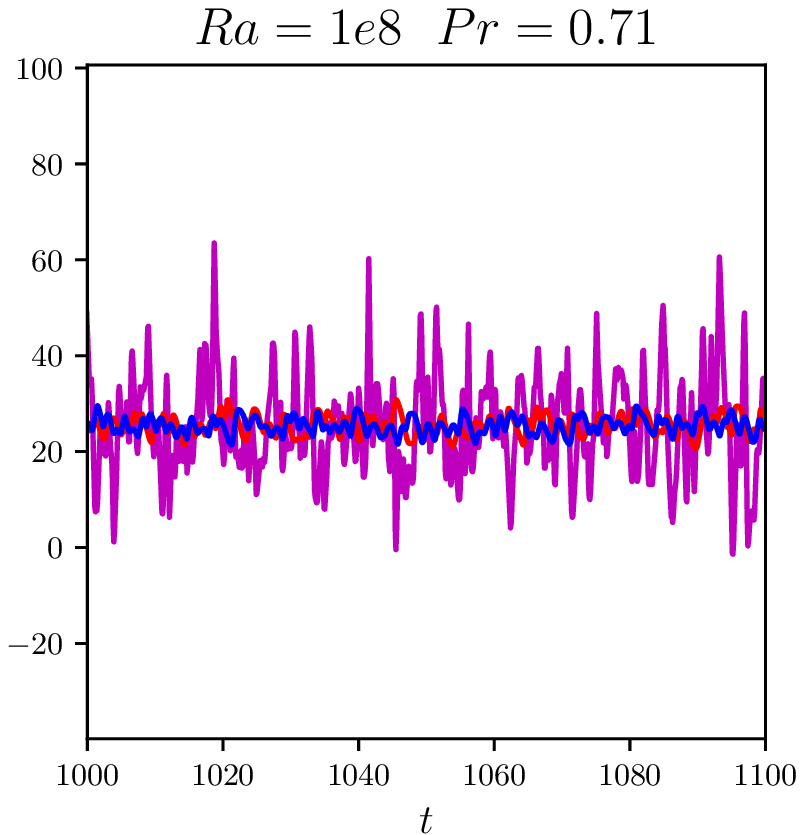}
		\includegraphics[scale=0.5]{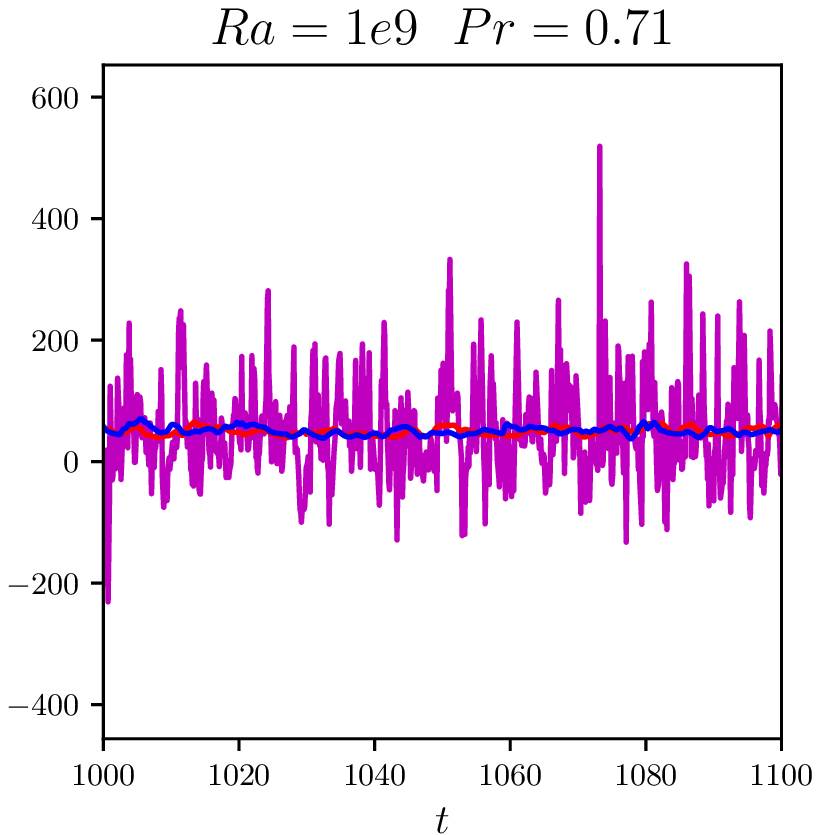}
		\includegraphics[scale=0.5]{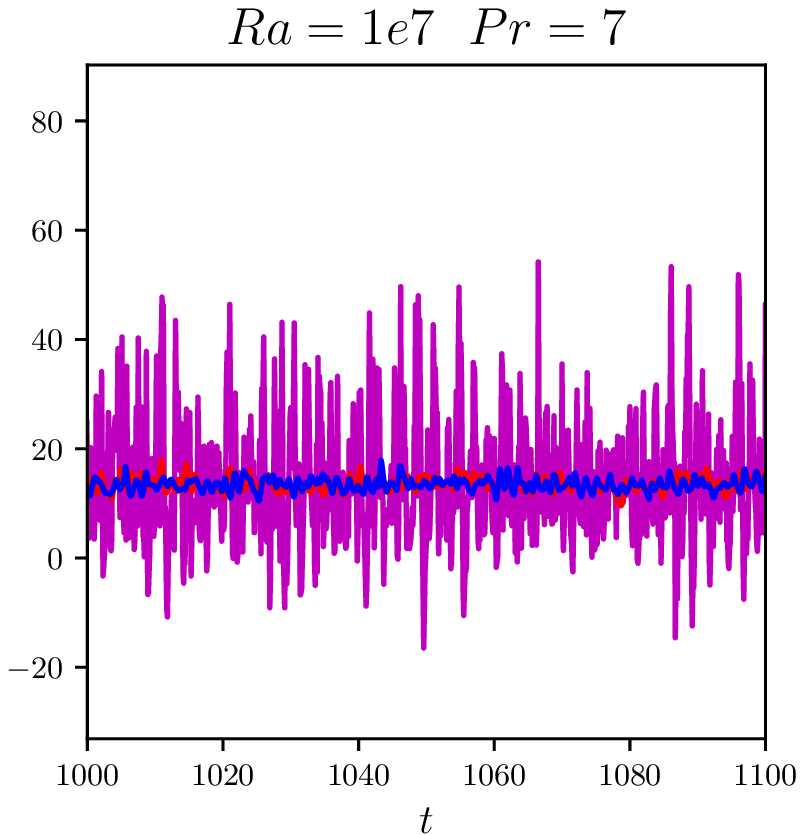}
		\includegraphics[scale=0.5]{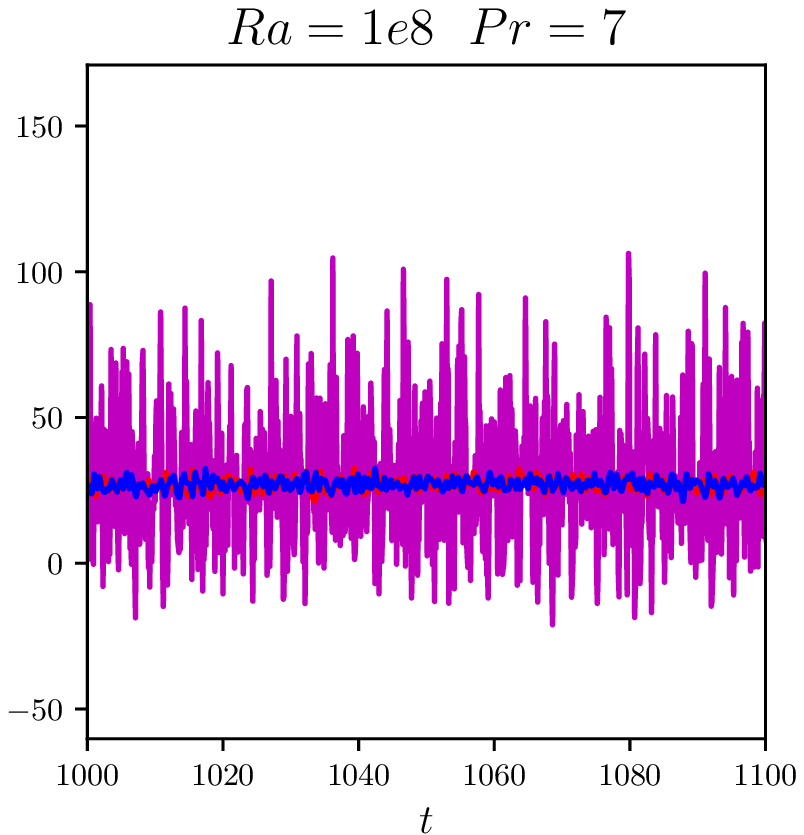}
		\includegraphics[scale=0.5]{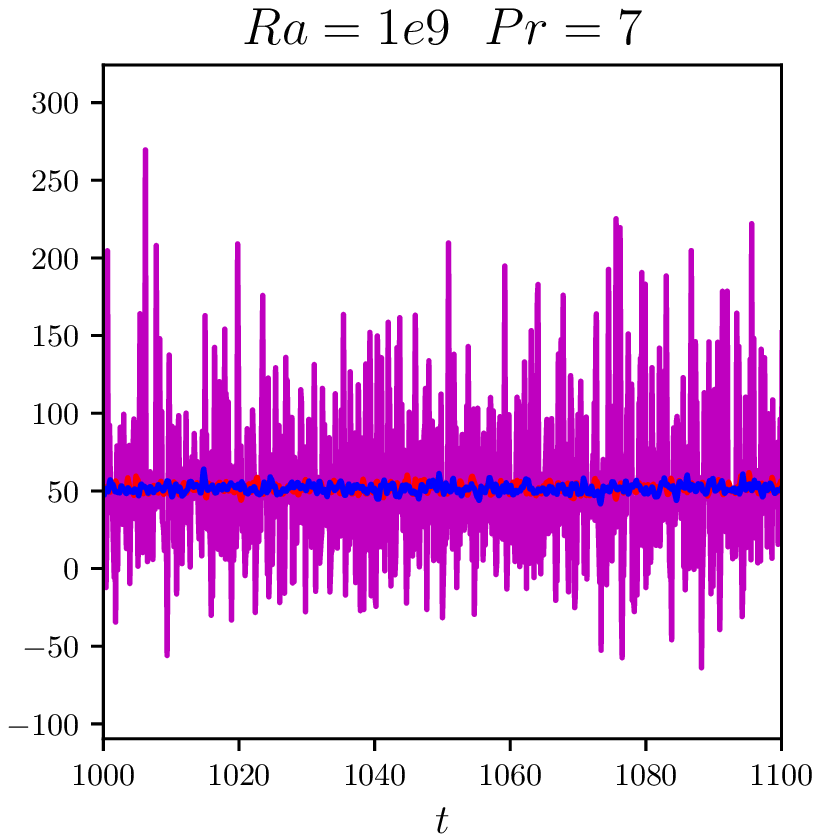}
		\caption{Direct recordings of $Nu_b$ (red), $Nu_m$ (magenta) and $Nu_t$ (blue) for $100$ eddy turnover times and various values of $Ra$ and $Pr$ in the statistically steady state. \label{Fig:Signals}}
	\end{center}
\end{figure}

To get some idea of the kind of flow displayed in the present configuration, in Fig. \ref{Fig:TemperatureVelocity} we show some instantaneous visualisation of the temperature and the velocity fields at different $Ra$, for the same Prandtl number ($Pr=0.71$).
It is known that in two dimensions there is a coherent angular flow up to moderately high Rayleigh numbers for moderate-Pr flows~\cite{sugiyama2010flow,podvin2015}. 
In particular, at low-Ra numbers $Ra<10^8$ it has been observed an intermittent behavior between coherent flow reversals and incoherent cessation periods~\cite{castillo2019cessation}, most notably at $Pr\sim4$.
Even though there is a dependence on the Prandtl number, the reversals and with them the coherent periods basically disappear for $Ra>10^8$.
With our configuration at $Pr=0.71$, reversals are basically absent as found in the past~\cite{sugiyama2010flow}, and yet there is a clear coherent flow at $Ra=10^8$, as nicely shown by \ref{Fig:TemperatureVelocity}(a),
where heat is mostly transported along this smooth ``wind'', as reflected by an almost bimodal velocity distribution displayed in Fig. \ref{Fig:TemperatureVelocity}c.
At higher $Ra$,  the flow becomes chaotic and coherence is lost, as highlighted in Fig. \ref{Fig:Signals}.
At $Pr=7$, the flow is much less organised and the flow loses coherence at a much lower Rayleigh number.

We can get some more insights about the dynamics of our system at looking at the temporal signals of the relevant quantities, namely the time-dynamics of the Nusselt numbers integrated over the homogeneous horizontal direction.
These signals are displayed in Figure \ref{Fig:Signals}, for the two Prandtl numbers investigated at different Rayleigh numbers.

The inspection of the signals indicates that for $Ra<10^8$ at $Pr=0.71$, the flow is dominated by the large-scale flow which displays a clear periodic behavior.
This is related to the periodic oscillations of the vortices located in the corners,
which can be seen in \ref{Fig:TemperatureVelocity}a and have been analysed in details in previous studies~\cite{castillo2016reversal,castillo2017turbulent}.

Because of these changes in direction the fluctuations are very large. The same periodicity is reflected by the signals recorded at the boundaries,  even though the friction is able to damp the amount of variation. 
As expected~\cite{sugiyama2010flow}, for the case at high Prandtl number the transition toward a fully chaotic state happens before and not much difference is found among flows at different $Ra$ numbers, from a qualitative point of view.
These observations are confirmed by looking at Nusselt correlations (shown in appendix \ref{app:corr}), which indicate that the periodicity has been lost at $Ra\approx 10^8$ for $Pr=0.71$, and already for $Ra<10^7$ at $Pr=7$.
A first important conclusion about our statistics is that the signals at the top and bottom walls are statistically similar despite the different temperatures, but as physically expected from the symmetry of the system. That allows us to consider only one of them in the following, and actually to use the other to improve statistics.
Then, we can observe as a general feature that fluctuations are much larger in the mid-plane than at the bottom and top boundaries, as quantitatively witnessed by the values of the standard deviation given in appendix \ref{Appendix:Nusselts}. 
This evidence is in line with the phenomenology of turbulence, as the core of the flow is basically inertial and less affected by viscosity, whereas the effect of viscosity is dominant at the boundaries. Still, it is interesting to remark that non-negligible fluctuations are produced also at the boundaries.
Generally speaking, the results of this section are meant to show that the present results are consistent with previous studies, and to highlight differences between low and high Prandtl number flows.

\subsection{Statistics of the heat-flux}
\label{section3b}

\begin{figure}[h]
	\begin{center}
		\includegraphics[scale=0.5]{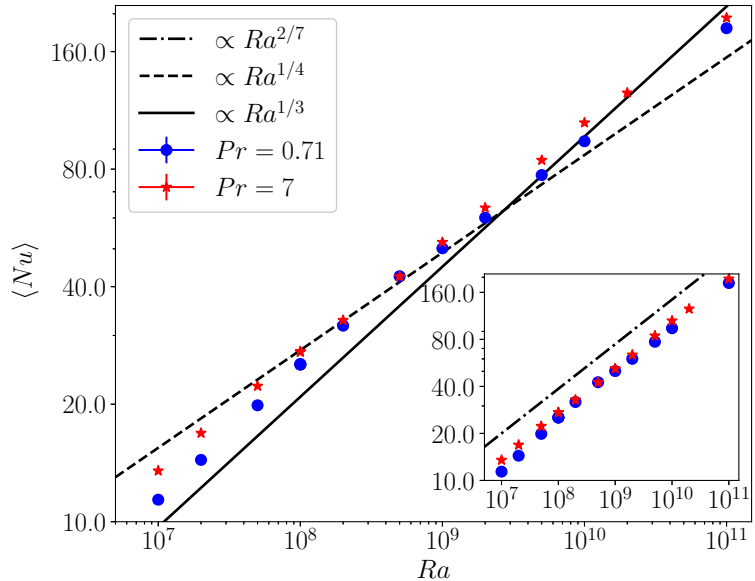}
		\caption{Mean Nusselt number as a function of $Ra$. 
		The scaling laws $\langle Nu \rangle \propto Ra^{1/4}$, which corresponds to a laminar boundary layer, and $\langle Nu \rangle \propto Ra^{1/3}$, which corresponds to the Malkus scaling \cite{malkus1954heat,malkus1954discrete}, and are plotted in the main graph to guide the eye.
		In the inset, also the scaling $\langle Nu \rangle \propto Ra^{2/7}$ is plotted~\cite{castaing1989scaling}.}
		\label{Fig:AverageNusselt}
	\end{center}
\end{figure}
\begin{figure}[h]
	\begin{center}
		\includegraphics[width=0.9\textwidth]{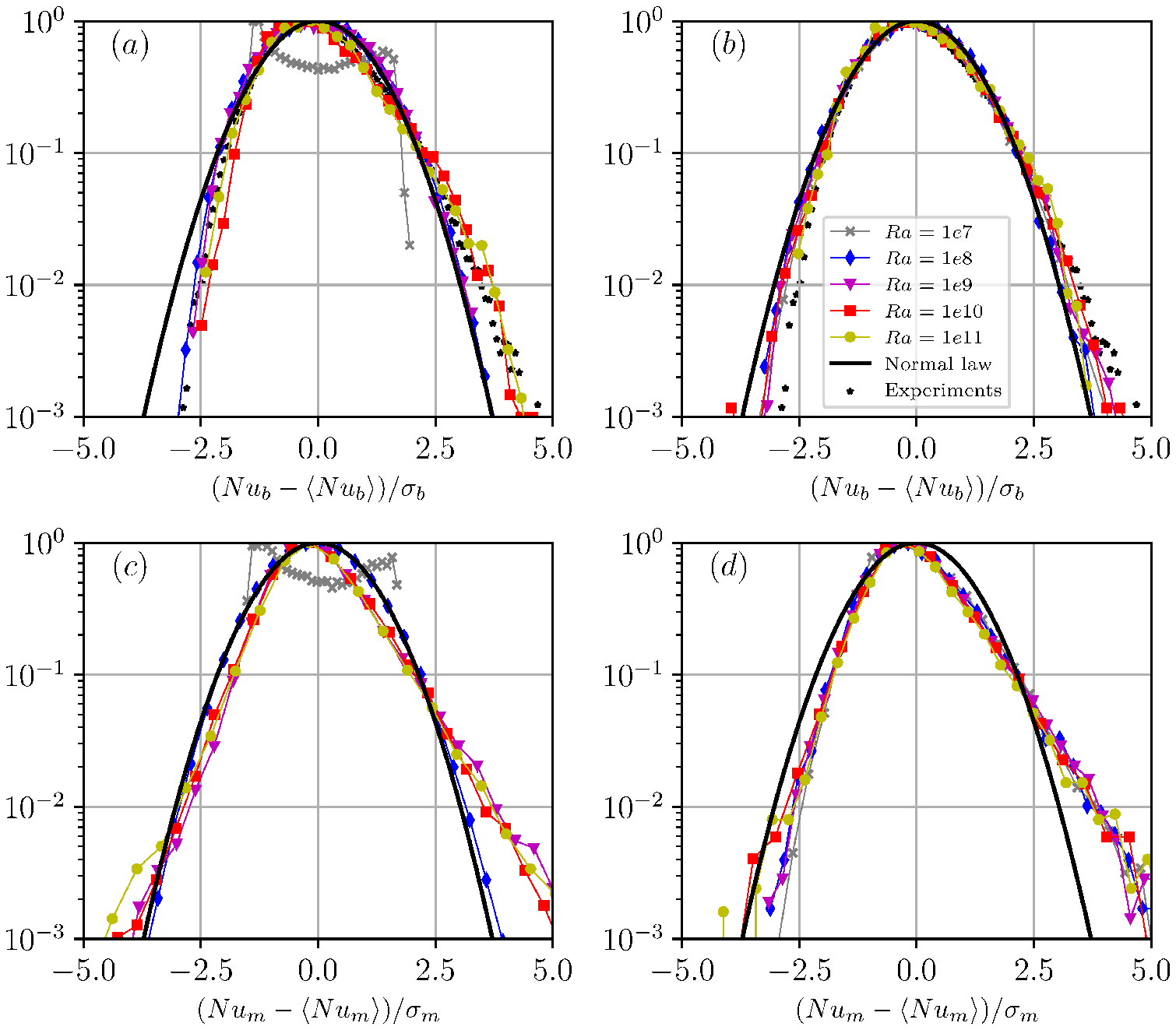}
		\caption{Pdfs of the reduced Nusselt numbers at the bottom boundary and at the middle plane for various $Ra$ and $Pr$. (a) Bottom boundary for $Pr=0.71$. (b) Bottom boundary for $Pr=7$. (c) Middle line for $Pr=0.71$. (d) Middle line for $Pr=7$. 
		The experimental data are shown for $Ra=1.7e9$, $Pr=7$ and $\Gamma=1/2$ only at the bottom walls, since they are not available elsewhere.
		A Gaussian curve is also plotted to help the comparison. 
		\label{Fig:CenteredPDF}}
	\end{center}
\end{figure}
In figure \ref{Fig:AverageNusselt}, we show the plot of the mean Nusselt number against the $Ra$ number for our simulations. 
Results might support the existence of two regimes in the parameter space investigated here, and differences between the two $Pr$ are recorded.
At $Pr=7$, the laminar scaling appears to be found in the range
$ 5 \times 10 ^7 \lesssim Ra \lesssim 10^9$, and the 
$1/3$ scaling is retrieved for $  10^9 \lesssim Ra \lesssim 2\times10^{10}$. Points at lower $Ra$ are possible slightly different, while the simulation at highest $Ra$ is slightly under the power law, which might suggest the beginning of a new scaling or be just an effect of statistical error. 
Results at lower $Pr$ might suggest instead a laminar scaling in the range $ 5 \times 10^8 \lesssim Ra \lesssim 10^{10}$. 
For lower $Ra$, while we have verified that the flux related to each modes is comparable, its value is yet impacted by the periodic structure of the flow and the heat-flux is damped.
For higher $Ra$ a transition to the $1/3$ scaling may be also observed, but the present results do not allow to draw a neat conclusion. 
Results are consistent with the physical picture provided above, with the results obtained in previous studies~\cite{zhang2017statistics, van2012flow}, and also with the experimental results considered here~\cite{aumaitre2003statistical}.
What is important to emphasize is that the scaling law is difficult to precisely determine, as highlighted by the inset figure where an intermediate scaling $2/7$, proposed on some physical ground~\cite{castaing1989scaling}, appears to fit the data correctly over 4 decades for both $Pr$.

In figure \ref{Fig:CenteredPDF}, we show the pdf of the reduced heat-flux at the boundaries and in the mid-plane of the domain, $(Nu_b - \langle Nu_b \rangle) / \sigma_b$ and $(Nu_m - \langle Nu_m \rangle) / \sigma_m$, at  different $Ra$ numbers,  for the two Prandtl numbers $Pr=0.71$ of $Pr=7$. 
We compare the results with experiments \cite{aumaitre2003statistical}, for which it was found that the pdf at the boundaries was almost independent of $Ra$ and the aspect ratio, and slightly skewed with respect to a Gaussian.

For $Pr=0.71$ (left column), important changes are observed when increasing the Rayleigh number.
 In particular, the pdf of the reduced Nusselt goes from a bimodal distribution at $Ra=1e7$ to a skewed distribution for $Ra \gtrsim 10^{11}$. 
 The same behavior is visible at the walls and at the middle of the domain, even in a more marked way.
 Results are consistent with the fact that in 2D at low Prandtl numbers the flow is basically periodic at low $Ra$ numbers,  while for high $Pr$ the chaotic state has  been already attained at moderate $Ra$, as highlighted by the temporal signals in Fig. \ref{Fig:Signals}.
Moreover, the pdf is almost Gaussian for $Ra < 10^9$ at the mid-plane, while it is non-Gaussian and skewed for higher $Ra$.
We observe, notably, that the pdfs do not change anymore for $Ra\ge 10^{10}$.
That might indicates a transition 
around a value $Ra \approx 10^9\div10^{10}$.

As expected, at $Pr=7$ (right column) no bimodal pdf is found at all, confirming that the transition to a chaotic state has been reached even for the lowest $Ra$.
Furthermore, at variance with the low-Pr case, the pdf is found to be quasi-normal at the boundaries, for $Ra\lesssim 10^8$. 
Then, the pdf starts to be more and more skewed, 
suggesting a later transition between $Ra\sim 10^{10}$ and $Ra\sim 10^{11}$.
At the mid-plane, the pdfs are  instead similar for all $Ra$ and are all well skewed, with only some possible differences in the negative tail at the middle of the domain, where nonetheless statistical errors may affect the results. Thus, the pdfs in the mid-plane do not permit to capture any transition, and suggest a turbulent state in the core of the domain at all Ra.

Globally speaking, some differences are displayed by fluctuation profiles at the bottom boundary and in the mid-plane, with stronger fluctuations in the bulk, as manifested by the larger tails.
The results are found to be in quite good agreement with the experiments, where such pdfs were recorded only at the boundaries.
As expected, the comparison is nicer for the numerical results obtained for $Pr=7$, although the results for $Pr=0.71$ are not much different for $Ra>10^9$.


\subsection{Scaling laws for the ratio of the fluctuations to the mean value of the heat flux}
\label{section3c}
\begin{figure}[h]
	\begin{center}
		\includegraphics{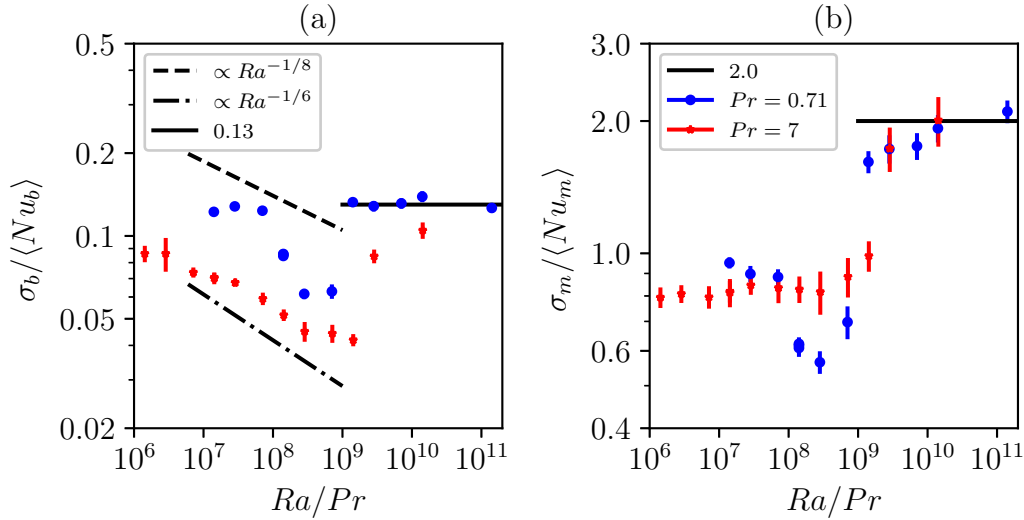}
		\caption{Relative rms fluctuations (rms fluctuations divided by the mean value) of the Nusselt number Eqs. (\ref{eq:sigma}): (a) at the boundaries; (b) in the mid-plane.}
		\label{Fig:RMS}
	\end{center}
\end{figure}
\begin{figure}[h]
	\begin{center}
		\includegraphics{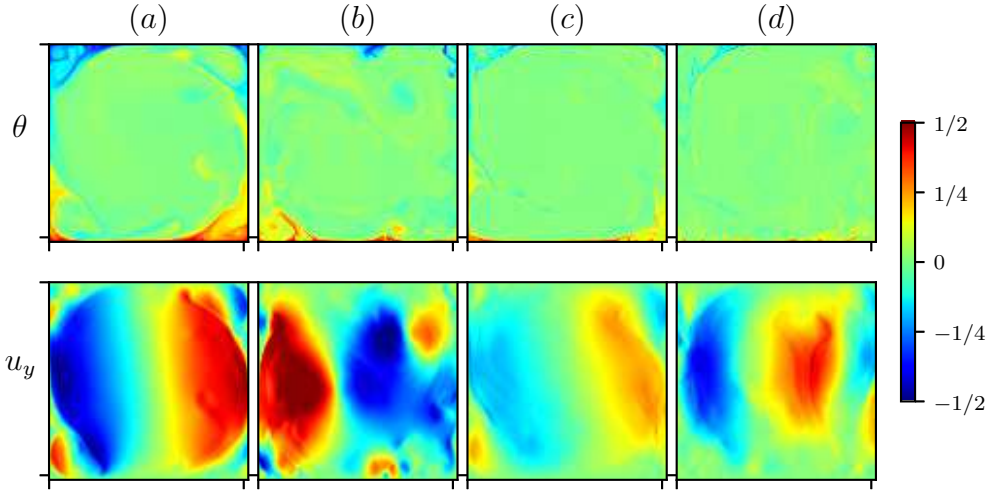}
		\caption{Temperature $\theta$ and vertical velocity $u_y$ snapshots in statistically steady state for various $Ra$ and $Pr$: (a) $Ra=5 \, 10^8$ and $Pr=0.71$, (b) $Ra=2 \, 10^9$ and $Pr=0.71$, (c) $Ra=5 \, 10^9$ and $Pr=7$, (d) $Ra=2 \, 1010$ and $Pr=7$. We observe the turbulent transition of the kinetic boundary layer around $Ra/Pr \simeq 10^9$ for both $Pr=0.71$ and $Pr=7$. 
		}
		\label{Fig:TurbulentTransition}
	\end{center}
\end{figure}

In previous experiments~\cite{aumaitre2003statistical}, it has been observed that the root mean square of the heat-flux at walls $\sigma_b$ divided by the mean heat-flux displays a power law on almost three decades in Rayleigh number for $10^6 < Ra/Pr < 10^9$ . 
The proposed rationale behind this behavior is the following:
if one takes the thermal boundary layer thickness $\delta_T$ as the relevant correlation length along the wall, which is the surface of measurement, then it can be considered that the average heat-flux measured over an area $H^2$ consists of the sum of $N=(H/\delta_T)^2$ uncorrelated contributions. 
Using the law of large numbers, the rms of the heat-flux should therefore scale as $\propto \sqrt{N}=H/\delta_T$. 
We adapt here the same argument to two dimensions, such that the average heat-flux is computed over a line of length $H$ so it consists of the sum of $N=H/\delta_T$ uncorrelated contributions. Consequently, the rms of the heat-flux should scale as $\propto \sqrt{H/\delta_T}$. Using the relation between the Nusselt number and the thermal boundary layer thickness $\langle Nu_b \rangle= H/2\delta_T$, we obtain the following scaling law:
\begin{equation}
\frac{\sigma_b}{\langle Nu_b \rangle} \propto \frac{\sqrt{H/\delta_T}}{H/2 \delta_T} \propto \frac{1}{\sqrt{\langle Nu_b \rangle}}~.
\label{eq:Aumaitre}
\end{equation}
This relation is plotted in Fig. \ref{Fig:RMS} for our 2D simulations as a function of $Ra/Pr$. 
Since the scaling exponent of the average Nusselt number has been found to be between $1/4$ and $1/3$, we have represented the lines $\propto Ra^{-1/8}$ and $\propto Ra^{-1/6}$ to check the validity of the  argument given by equation (\ref{eq:Aumaitre}).
 
Let us consider first the walls, for which we have also experimental data, see \ref{Fig:RMS}(a).
For $Pr=7$ and $Ra/Pr \leq 10^9$, that corresponds to the experimental range, the dependance of $\sigma_b /\langle Nu_b \rangle$  is compatible with the scaling relation (\ref{eq:Aumaitre}).
However, increasing $Ra$ we find a clear transition with this ratio reaching an almost constant value above $Ra/Pr = 2\times 10^9$. 

For $Pr=0.71$, we also find a decrease of the relative fluctuations with increasing Rayleigh number up to $Ra/Pr \gtrsim10^9$ followed by a regime where $\sigma_b / \langle Nu_b \rangle$ reaches a constant value $\simeq 0.13$. However, no scaling law is observed below $10^9 \gtrsim Ra/Pr$ because of the transition from a nearly periodic flow to a chaotic flow that occurs within that range.

Let us analyse the behavior of the heat-flux rms fluctuations in the mid-plane, shown in Fig. \ref{Fig:RMS}b. 
Remarkably, a transition is also observed, basically at the same $Ra/Pr$ for both $Pr$ as encountered at the boundaries.
However, the power law predicted at low $Ra$ at the boundaries is not found, instead  the relative fluctuations of the heat-flux take constant values both below and above the transition, and independently of $Pr$. At the transition, the relative fluctuations increase abruptly by more than a factor two.
Moreover, the amplitude of $\sigma_m$ is  more than one order of magnitude higher than the one at the boundaries.  

We understand the presence of a plateau following a reasoning similar to that used to rationalise the dissipation  anomaly of $\epsilon$ \cite{frisch1995turbulence}.
In the core of the flow, the boundary layer does not play any role so that the correlation length should be of the order of the integral length, which turns out to be of the order of $L_{\text{int}}\sim 0.2 H$ in the core of the flow with little dependence on Rayleigh, at least for $Ra\lesssim 10^9$~\cite{kaczorowski2013turbulent}.
Moreover, it has been ingeniously observed~\cite{cadot1997energy} that the scaling of dissipation rate at walls depends on the viscous boundary scale when it is stable; yet,  if the kinetic boundary layer is destabilised, a turbulent transition is induced and the dissipation rate reaches a plateau also at walls, though with a smaller prefactor than in the bulk.
Here, the scenario appears similar. At the boundaries, a stable viscous boundary layer induces the power-law scaling (\ref{eq:Aumaitre}), whereas a plateau is found when the kinetic boundary layer becomes unstable. The prefactor is about one order of magnitude less than the corresponding one computed in the mid-plane. 
The same transition seems to induce also an increases of the fluctuations, measured by a large prefactor, in the mid-plane. About this behavior, we can only guess that it is related to the presence of strongly intermittent plumes.

The transition at $Ra/Pr \sim10^9$ can be traced back to the destabilization of the  kinetic boundary layer.
To give further evidence of that, we display in Fig. \ref{Fig:TurbulentTransition} the temperature and velocity fields in the vicinity of the transition. It is clear that turbulent spots in the velocity field appear only after the transition. Moreover, at high Prandtl number the temperature field does not display significant differences, being already fully turbulent, whereas at $Pr=0.71$ the temperature also shows an intermittent behavior only after the transition.
As highlighted by the Fig. \ref{Fig:TurbulentTransition}, the transition changes the global dynamics of the flow, strongly increasing the intermittent behavior and hence the probability of strong events.
This is why the transition is experienced for the same value of the Rayleigh number also in the mid-plane, although the dynamics there is basically inviscid and the boundary layer has no direct influence. Indeed, the increase of the constant through the transition indicates a sudden increase of the fluctuations.

\section{Discussion and conclusions\label{section4}}

In this work, we have performed a statistical analysis of the fluctuations of the time-dependent heat-flux integrated over the horizontal direction, at the bottom and top boundaries and in the mid-plane, in turbulent Rayleigh-Bénard convection.
The heat-flux is indeed averaged over a small spatial region, and constitutes a random variable.
This idea to look at such fluctuations was pioneered in a previous experimental study~\cite{aumaitre2003statistical}, and we have complemented it through extensive direct numerical simulations in a 2D geometry. In fact, statistics of heat-flux fluctuations have been rarely considered previously.

The simplicity of the 2D configuration with respect to the 3D one has permitted us to obtain results over 4 decades in Rayleigh number for two Prandtl number flows, and to assure a good statistical convergence both for the pdfs and the second moments, which is key for the kind of statistical analysis carried out here.
Moreover, numerical simulations have allowed to access new information, notably about the fluctuations inside the volume.
The comparison of the pdfs given by the experiments with those obtained here numerically on a larger range of Rayleigh numbers has shown that while some quantitative differences are found between 2D and 3D results, the qualitative behavior is similar especially for the cases with the same fluid properties. 
The only significant difference is at low Rayleigh numbers, $Ra \lesssim 10^8$, where the 2D dynamics is dominated by quasi-periodic patterns and a bimodal pdf is displayed that is not observed in 3D.
The quite good agreement of the pdfs between the 2D and the 3D data shows that the results obtained in 2D should be relevant also for the 3D realistic case.


The main finding of the present study is to point out that the scaling of the root-mean-square of the heat-flux with respect to the ratio 
$Ra/Pr$ displays a bifurcation at $Ra/Pr \sim 10^9$.
This has been observed for two different Prandtl numbers and both at the bottom and top boundaries and in the bulk of the flow.
In particular, at the boundaries the transition is between a regime where $\sigma/\lra{Nu}$ decreases with $Ra$, toward one where this ratio is constant.

This clear transition was not observed in the previous experiments because the critical value of  $Ra/Pr $ was not reached at the time.
Besides, no other work has analysed such statistical scaling.

An important point is that the standard analysis of the scaling of  the mean Nusselt number with respect to $Ra$ could be qualitatively compatible with such a transition, but does not clearly point it out. 
The transition might be related to the change of slope between the laminar scaling $Nu \sim Ra^{1/4}$ to the turbulent one $Nu \sim Ra^{1/3}$, but one observes that in the range of $Ra$ of the present study, all the data are compatible with a scaling of the kind $Nu \sim Ra^{2/7}$, and therefore do not display any transition.
This is also consistent with what has been observed by previous authors~\cite{van2012flow,zhang2017statistics,zhang2017dissipations}.
Based on a series of 2D numerical simulations very similar to those presented here~\cite{zhang2017dissipations}, one of the conclusions was that a single scaling  $Nu \sim Ra^{0.3}$ (intermediate between $2/7$ and $1/3$) was compatible with the whole range, displaying no transition. A fresh look at the data might suggest eventually a mild change of slope at about $Ra/Pr \sim 10^9$, but data are not sufficient to claim that.

Another interesting finding is that even the pdfs of the heat-flux are less informative that the scaling of the variance.
Indeed, the pdfs are  compatible with a transition 
around $Ra/Pr \sim 10^9$, 
however the observation is based on the tails of the pdfs and is not clear-cut.
Interestingly, in another study~\cite{zhang2017dissipations}, the pdfs of the dissipation were displayed and a change in the tails was clearly observed also at $Ra/Pr \sim 10^9$. In a related work \cite{zhang2017statistics}, the authors also remarked that the turbulent energy production averaged over the whole cell is negative except for the highest Rayleigh number $Ra = 10^{10}$ at $Pr = 5.3$.

Therefore, the key result is the following: while looking at the scaling of the mean flux it is not possible to single out a clear transition since the changes in slope are at best of few percent, whereas the scaling of the variance of the heat-flux reveals it neatly with an abrupt jump of a factor larger than $2$.

The visual inspection of the velocity fields shows that turbulent spots chaotically released from boundaries are observed only after the transition. This should not be confused with the possible transition to the ultimate state when both kinetic and thermal boundary layers are turbulent, because that should occur at a much higher value of $Ra$. This breakup of the large-scale circulation has already been observed and discussed in earlier studies \cite{van2012flow, van2013comparison} for different aspect ratios.

Another interesting result of the present work is related to the behavior of the fluctuations below the transition which is not the same at the boundaries and in the bulk of the flow.
At the boundaries, for $Pr=7$ we observe a regime compatible with a scaling argument based on the law of the large numbers, as previously found in experiments~\cite{aumaitre2003statistical}.
The normalised rms decreases also at $Pr=0.71$ but with a less clear scaling, since in this case convection is expected to depend more strongly on the global circulation which induces large scales correlations. 
This is consistent with the fact that we observe higher decorrelation times at lower Pr. 

Both for $Pr=0.71$ and $Pr=7$, the relative fluctuation of the Nusselt numbers at the boundaries and in the mid-plane reach constant values above the transition. These values are different for the mid-plane and the boundaries, probably because there is no thermal fluctuations at top and bottom boundaries. 
In the core of the flow, dominated by inertial dynamics, the relation $\sigma \sim \lra{Nu}$ appears to be always valid, but with a prefactor that increases through the transition due to an increase of turbulent fluctuations.

In conclusion, the analysis of heat-flux fluctuations in an idealised case, has showed that these fluctuations are of paramount importance to understand the underlying dynamics of the flow and most notably the presence of possible bifurcations between different turbulent regimes. 
Although not investigated here, these results seem promising for future experiments at higher Rayleigh numbers, notably with regard to the ultimate-state transition.
These results complement recent significant studies, which have included some modelling of turbulent fluctuations to analyse the transition of the boundary layers and have found that the scaling of such mean properties like the Reynolds number is largely insensitive to the fluctuations.

We plan in a future experimental study to investigate the presence of this transition in 3D and to analyse more in detail the mechanisms underlying it.

\section*{Aknowlegements}

This work was granted access to the HPC resources of the MeSU platform at Sorbonne-Université.

\appendix

\section{Numerical details \label{Appendix:Nusselts}}
\begin{table*}
	\begin{center}
	\small
		\begin{tabular}{@{}|c|c|c|c|c|c|c|c|}
			\hline
			\textbf{Run \#}  & $\boldsymbol{\langle Nu_b \rangle}$ & $\boldsymbol{\langle Nu_m \rangle}$ & $\boldsymbol{\langle Nu_t \rangle}$ & $\boldsymbol{\sigma_b}$ & $\boldsymbol{\sigma_m}$ & $\boldsymbol{\sigma_t}$ \\
			\hline
			1 & 11.39 $\pm$ 0.02 & 11.39 $\pm$ 0.15 & 11.39 $\pm$ 0.02 & 1.39 $\pm$ 0.03 & 10.84 $\pm$ 0.24 & 1.39 $\pm$ 0.03 \\ 
			2 & 14.40 $\pm$ 0.03 & 14.41 $\pm$ 0.22 & 14.41 $\pm$ 0.03 & 1.84 $\pm$ 0.06 & 12.94 $\pm$ 0.45 & 1.85 $\pm$ 0.06 \\ 
			3 & 19.87 $\pm$ 0.04 & 19.88 $\pm$ 0.25 & 19.88 $\pm$ 0.03 & 2.45 $\pm$ 0.08 & 17.56 $\pm$ 0.61 & 2.38 $\pm$ 0.08 \\ 
			4 & 25.25 $\pm$ 0.01 & 25.26 $\pm$ 0.02 & 25.25 $\pm$ 0.01 & 2.17 $\pm$ 0.01 & 15.72 $\pm$ 0.06 & 2.18 $\pm$ 0.01 \\ 
			5 & 25.35 $\pm$ 0.03 & 25.36 $\pm$ 0.21 & 25.38 $\pm$ 0.03 & 2.14 $\pm$ 0.08 & 15.66 $\pm$ 0.61 & 2.17 $\pm$ 0.08 \\ 
			6 & 25.28 $\pm$ 0.03 & 25.31 $\pm$ 0.22 & 25.30 $\pm$ 0.03 & 2.17 $\pm$ 0.08 & 15.42 $\pm$ 0.64 & 2.20 $\pm$ 0.08 \\ 
			7 & 31.78 $\pm$ 0.03 & 31.75 $\pm$ 0.25 & 31.72 $\pm$ 0.03 & 1.96 $\pm$ 0.08 & 17.95 $\pm$ 0.99 & 1.96 $\pm$ 0.09 \\ 
			8 & 42.53 $\pm$ 0.05 & 42.51 $\pm$ 0.54 & 42.55 $\pm$ 0.05 & 2.67 $\pm$ 0.16 & 29.66 $\pm$ 2.35 & 2.52 $\pm$ 0.15 \\ 
			9 & 50.16 $\pm$ 0.08 & 50.29 $\pm$ 0.94 & 50.19 $\pm$ 0.08 & 6.65 $\pm$ 0.19 & 81.20 $\pm$ 3.93 & 6.61 $\pm$ 0.19 \\ 
			10 & 60.09 $\pm$ 0.12 & 60.08 $\pm$ 1.61 & 59.90 $\pm$ 0.12 & 7.71 $\pm$ 0.33 & 103.84 $\pm$ 6.29 & 7.85 $\pm$ 0.33 \\ 
			11 & 77.26 $\pm$ 0.14 & 77.39 $\pm$ 1.92 & 77.53 $\pm$ 0.14 & 10.15 $\pm$ 0.43 & 135.75 $\pm$ 7.83 & 10.12 $\pm$ 0.39 \\ 
			12 & 94.31 $\pm$ 0.19 & 94.48 $\pm$ 2.57 & 93.95 $\pm$ 0.17 & 13.11 $\pm$ 0.59 & 181.92 $\pm$ 10.57 & 11.83 $\pm$ 0.51 \\ 
			13 & 183.70 $\pm$ 0.25 & 182.56 $\pm$ 4.10 & 183.82 $\pm$ 0.24 & 23.22 $\pm$ 0.91 & 383.78 $\pm$ 18.50 & 22.79 $\pm$ 0.78 \\ 
			14 & 13.50 $\pm$ 0.01 & 13.52 $\pm$ 0.14 & 13.50 $\pm$ 0.01 & 1.16 $\pm$ 0.08 & 10.73 $\pm$ 0.52 & 1.17 $\pm$ 0.09 \\ 
			15 & 16.87 $\pm$ 0.02 & 16.91 $\pm$ 0.16 & 16.87 $\pm$ 0.02 & 1.45 $\pm$ 0.20 & 13.68 $\pm$ 0.57 & 1.41 $\pm$ 0.20 \\ 
			16 & 22.26 $\pm$ 0.02 & 22.24 $\pm$ 0.26 & 22.26 $\pm$ 0.02 & 1.64 $\pm$ 0.07 & 17.68 $\pm$ 0.93 & 1.59 $\pm$ 0.07 \\ 
			17 & 27.31 $\pm$ 0.03 & 27.46 $\pm$ 0.39 & 27.32 $\pm$ 0.03 & 1.92 $\pm$ 0.09 & 22.36 $\pm$ 1.49 & 1.91 $\pm$ 0.09 \\ 
			18 & 27.20 $\pm$ 0.03 & 27.11 $\pm$ 0.31 & 27.20 $\pm$ 0.03 & 1.92 $\pm$ 0.08 & 22.20 $\pm$ 1.13 & 1.94 $\pm$ 0.07 \\ 
			19 & 32.83 $\pm$ 0.03 & 32.84 $\pm$ 0.32 & 32.82 $\pm$ 0.03 & 2.22 $\pm$ 0.07 & 27.80 $\pm$ 1.23 & 2.26 $\pm$ 0.07 \\ 
			20 & 42.46 $\pm$ 0.04 & 42.62 $\pm$ 0.63 & 42.47 $\pm$ 0.04 & 2.51 $\pm$ 0.13 & 35.56 $\pm$ 2.47 & 2.53 $\pm$ 0.13 \\ 
			21 & 51.99 $\pm$ 0.04 & 51.81 $\pm$ 0.70 & 52.00 $\pm$ 0.04 & 2.68 $\pm$ 0.13 & 42.92 $\pm$ 2.68 & 2.61 $\pm$ 0.13 \\ 
			22 & 63.68 $\pm$ 0.08 & 63.74 $\pm$ 1.37 & 63.71 $\pm$ 0.08 & 2.86 $\pm$ 0.23 & 52.08 $\pm$ 5.31 & 2.91 $\pm$ 0.24 \\ 
			23 & 84.38 $\pm$ 0.08 & 85.00 $\pm$ 1.65 & 84.44 $\pm$ 0.08 & 3.73 $\pm$ 0.28 & 75.20 $\pm$ 7.02 & 3.68 $\pm$ 0.24 \\ 
			24 & 105.20 $\pm$ 0.07 & 105.16 $\pm$ 1.74 & 105.23 $\pm$ 0.07 & 4.40 $\pm$ 0.22 & 103.74 $\pm$ 7.39 & 4.47 $\pm$ 0.22 \\ 
			25 & 125.46 $\pm$ 0.24 & 125.15 $\pm$ 4.85 & 125.32 $\pm$ 0.24 & 10.59 $\pm$ 0.60 & 216.79 $\pm$ 20.90 & 10.68 $\pm$ 0.64 \\ 
			26 & 195.28 $\pm$ 0.49 & 194.38 $\pm$ 9.33 & 194.03 $\pm$ 0.49 & 20.46 $\pm$ 1.38 & 390.33 $\pm$ 41.05 & 20.45 $\pm$ 1.42 \\ 
			\hline
		\end{tabular}
		\caption{ Average and rms of the Nusselt numbers at boundaries and middle line. Their statistical errors are estimated supposing a decorrelation time of $2.0$ convective timescale.}
		\label{tab2}
	\end{center}
\end{table*}
In Table \ref{tab2} we show the mean Nusselt number computed for each 
case at the three positions considered. 
The corresponding statistical error is given by the standard deviation $\sigma_i$, for which we also indicate the statistical error.
We can see that in all cases the mean Nusselt numbers computed at different height are fully consistent, confirming that these simulations attained a statistically steady state.
The statistical errors have been used as error bars in figures. 
In fact, errors might be slightly larger because of possible residual correlations, notably concerning the error on $\sigma$.

We have further verified the grid convergence of our simulations at $Ra = 1e8$ and $Pr=0.71$ also by computing $\langle Nu_b \rangle$, $\langle Nu_m \rangle$, $\langle Nu_t \rangle$, $\sigma_b$, $\sigma_m$ and $\sigma_t$ for different mesh resolutions (see runs $4,5$ and $6$ in tables \ref{tab1} and \ref{tab2}). 
For these simulations, the discrepancies are of less than $2\%$.

\section{Nusselt correlations}
\label{app:corr}
\begin{figure}[h]
	\begin{center}
				\includegraphics[scale=0.85]{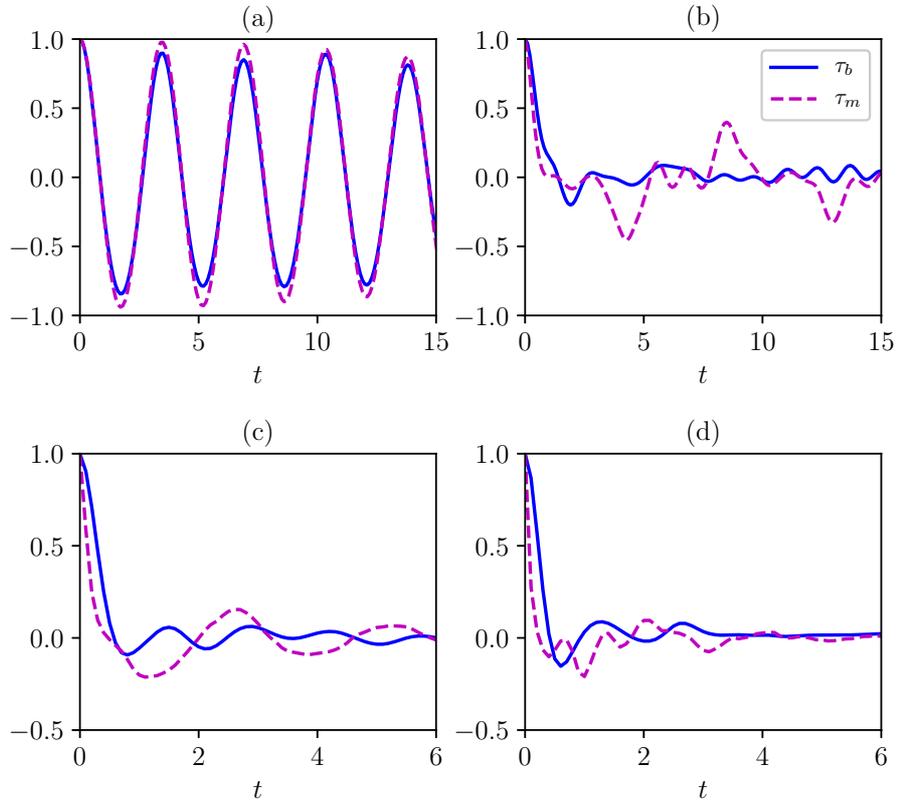}
		\caption{Autocorrelations of the Nusselt numbers at the bottom $\tau_b$ and at mid-plane $\tau_m$ defined as $\tau_\alpha=1/T \int_{t_0}^{t_0+T} Nu_\alpha(t_0) Nu_\alpha(t_0+t^\prime) dt^\prime
		$, with $\alpha=b,m$.
		(a) $Ra=10^7$ and $Pr=0.71$ , (b) $Ra=2e8$ and $Pr=0.71$ , (c) $Ra=1e7$ and $Pr=7$ , $Ra=2e8$ and (d) $Pr=7$. } \label{Fig:Correlation}.
	\end{center}
\end{figure}
To better appreciate the transition from the quasi-periodic state to a chaotic one, we show in Fig. \ref{Fig:Correlation} the auto-correlations of the signals $Nu_b$ and $Nu_m$ , respectively $\tau_b$ and $\tau_m$.
It is shown that at $Pr=0.71$ periodicity dominates the dynamics both at walls and in the core of the flow at $Ra\sim 10^7$.
The transition to a fully chaotic state is around $Ra\approx 2\times 10^8$, as indicated by Fig.\ref{Fig:Correlation}b, where the correlation rapidly goes to zero both at walls and in the middle. 
For the case at $Pr=7$ the periodicity has been already lost at $Ra = 10^7$.


\bibliographystyle{apalike}
\bibliography{biblio}

\end{document}